\documentclass[doublecol]{epl2} 
\usepackage{graphicx}
\usepackage{graphicx,amsfonts}
\usepackage{epsfig,amsmath}
\usepackage{verbatim}

\newcommand{\rmd}{{\mathrm{d}}}
\newcommand{\gras}[1]{\boldsymbol{#1}}

\title{Extremal statistics of curved growing interfaces in 1+1 dimensions}

\author{Joachim Rambeau\inst{1} \and Gr\'egory Schehr\inst{1}}
\shortauthor{J. Rambeau, G. Schehr}

\institute{
\inst{1} Laboratoire de Physique Th\'eorique (UMR du
  CNRS 8627), Universit\'e de Paris-Sud 11, 91405 Orsay Cedex,
  France
}


\abstract{
We study the joint probability distribution function (pdf) $P_t(M,X_M)$
of the maximum $M$ of the height and
its position $X_M$ of a curved growing interface belonging to the
universality class described by the Kardar-Parisi-Zhang equation in
$1+1$ dimensions, in the long time $t$ limit. We obtain exact results for the related problem of $p$ non-intersecting Brownian bridges
where we compute the joint pdf $P_p(M,\tau_M)$, for any finite $p$,  where $\tau_M$
is the time at which the maximal height $M$ is reached. This yields an
approximation of $P_t(M,X_M)$ for the interface problem, whose accuracy is systematically improved as $p$ is increased, becoming exact for $p \to \infty$. We show that our results, for moderate values of $p \sim 10$, describe accurately our
numerical data of a prototype of these systems, the polynuclear growth model in droplet geometry. We also discuss applications of our results 
to the ground state configuration of the directed polymer in a random medium with one
fixed endpoint. 
}

\begin{document}

\maketitle

\section{Introduction} The study of fluctuations in stochastic growth processes has attracted much attention
during the last two decades \cite{krug_book, halpin_review}. Such processes are ubiquitous in nature
as they appear in various physical situations ranging from paper wetting to burning fronts or growing bacterial colonies. In many experimental 
settings the growth starts from a point like region (a seed) with a strong tendency to evolve towards
the approximate spherical symmetry. Such examples include fluid flow in porous media, adatoms and vacancy islands on surfaces \cite{einstein} but also biological systems such as tumors~\cite{bru}. 

To describe such phenomena driven by a growing interface, several models have been studied, like the Eden model,  polynuclear growth models (PNG) or ballistic deposition models, among others \cite{halpin_review}. In $1+1$ dimensions, it is widely believed that all these models belong to the same
universality class as that of the Kardar-Parisi-Zhang (KPZ) equation
\cite{kpz,sasamoto_universality}. At time $t$, the width of the interface $W_L(t)$, for such 
a system of size $L$, behaves like $W_L(t) \sim L^{\zeta}{\cal W}(t/L^z)$ with universal exponents 
$\zeta = \frac{1}{2}$ and $z = \frac{3}{2}$ \cite{bethe}. In the
growth regime $t_{0} \ll t \ll L^z$ (where $t_{0}$ is a microscopic time scale) exact results
for different lattice models have shown that the notion of universality extends far beyond
the exponents and also applies to full distribution functions of physical observables \cite{praehofer_spohn, growth_exact, satya_review_tw}. In particular, the scaled cumulative distribution of the height field coincides with the Tracy-Widom (TW) distribution ${\cal F}_{\beta}(\xi)$ with $\beta = 2$ (respectively $\beta = 1$) for the curved geometry (respectively for the flat one), which describes the edge of the spectrum of random matrices in the Gaussian Unitary Ensemble (respectively of the Gaussian Orthogonal Ensemble) \cite{tracy_widom}. Height fluctuations were measured in experiments, both in planar \cite{exp_planar} and more recently in curved geometry in the electroconvection of nematic liquid crystals~\cite{exp_curved} and a very good agreement with TW distributions was found.

Here we focus on a prototype of these models, the PNG model \cite{png}, but 
our results hold more generally for curved growing interfaces in the KPZ universality class {(see below)}. It is defined as follows. At time $t=0$ a single island starts spreading on a flat substrate at the origin $x=0$ with unit velocity. Seeds of negligible size then nucleate randomly at a constant rate $\rho=2$ per unit length and unit time and then grow laterally also at unit velocity. When two islands on the same layer meet they coalesce. Meanwhile, nucleation continuously generates additional layers and in the droplet geometry, nucleations only occur above previously formed layers. Denoting by $h(x,t)$ the height of the interface at point $x$ and time $t$, one thus has 
$h(x,t) = 0$ for $|x| > t$. 
\begin{figure}[h]
 \onefigure[width= \linewidth]{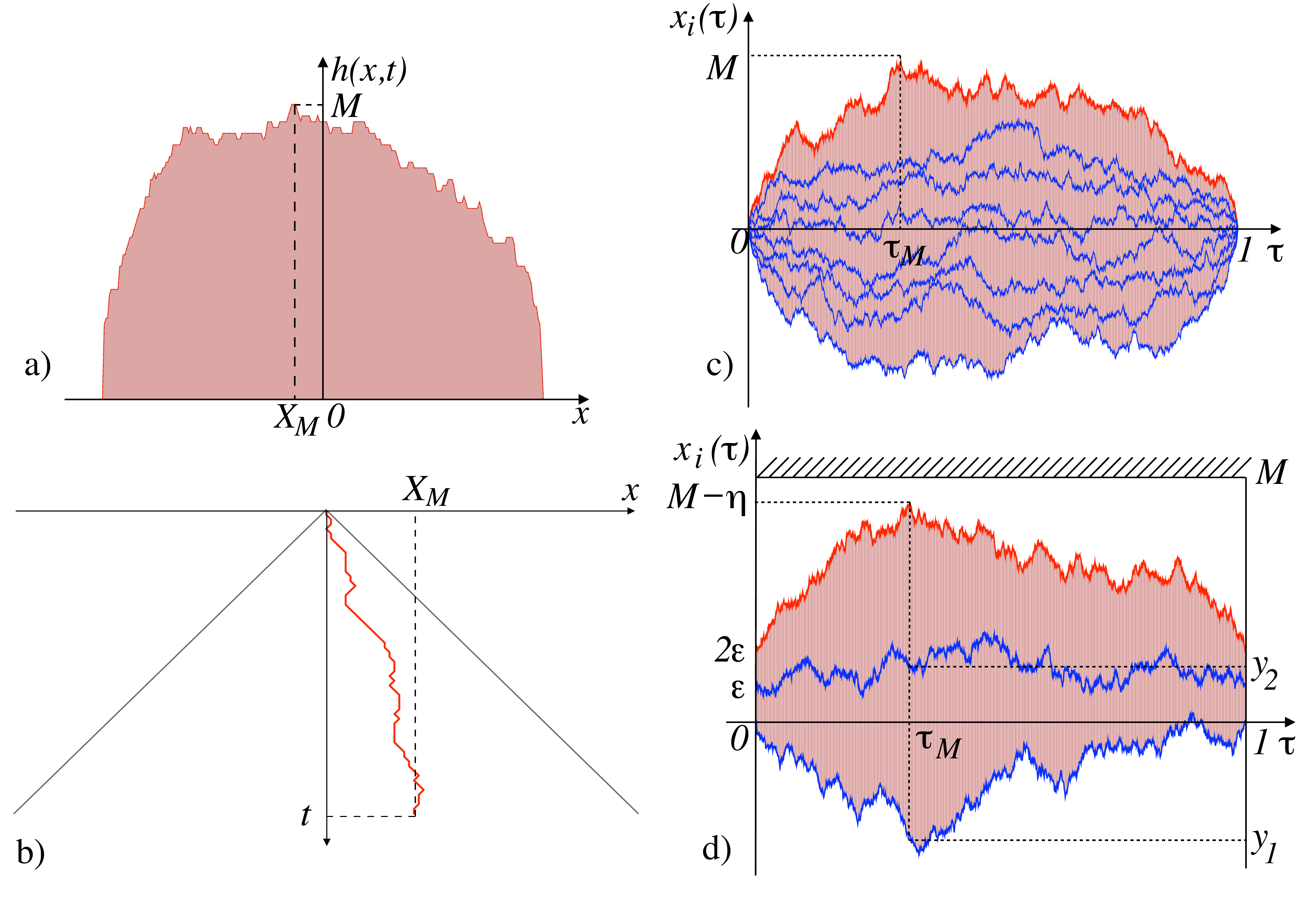}
\caption{{\bf a)}: Height profile $h(x,t)$ for fixed time $t$ for the PNG model.
{\bf b)}: An optimal path of the DPRM of length $t$. {\bf c)}: A $p$-watermelon configuration ($p=8$). {\bf d)}:~Sketch of the method to compute the joint pdf $P_p(M,\tau_M)$.}\label{fig_intro}
\end{figure}
On the other hand, in the long time limit, the profile for $|x| \leq t$ becomes circular $h(x,t) \sim 2t
\sqrt{1-(x/t)^2}$ \cite{praeho_thesis}, but there remain height
fluctuations around this semi-circle. A natural way
to characterize these fluctuations is to consider the maximal height
$M$ and its position $X_M$ (see Fig. \ref{fig_intro} a)). According to
KPZ scaling, one expects  $M - 2t\sim t^{\frac{1}{3}}$, while $X_M
\sim t^{\frac{2}{3}}$ \cite{johansson_transverse}.  The purpose of
this Letter is to provide an analytic approach to the joint pdf
$P_t(M, X_M)$ for large $t$.   

The PNG model
can be mapped onto the directed polymer in a random medium (DPRM) on
the square lattice with one fixed end \cite{halpin_review, satya_review_tw}. In
this language $M$ is the ground state energy while $X_M$ is the
transverse coordinate of 
the free end of the optimal polymer of length $t$ (see Fig. \ref{fig_intro} b)). Related questions for the continuum DPRM are currently
under active investigations \cite{dprm_doussal}. The marginal
distributions of $M$ and $X_M$ 
are already interesting and related extreme value quantities, like the
maximal relative height, have been extensively studied in the
stationary regime $t \gg L^z$~\cite{shapir, satya_airy, evs}. Much
less is known in the growth regime which we focus on. From the mapping
onto the DPRM, one identifies the pdf of the maximal height $M$
in the droplet geometry with the pdf of the height $h_{\rm flat}(x,t)$
at a given point $x$ and time $t$ in the planar geometry
\cite{krug_dprm}. Therefore we conclude that the pdf of $M$, suitably
rescaled and shifted, is given by ${\cal F}'_1(\xi)$ \cite{baik_rains,
  johansson_transverse}, the TW distribution for $\beta = 1$. On the other hand, the computation of the (marginal) distribution of $X_M$
is a challenging open problem \cite{johansson_transverse}.  

To compute $P_t(M,X_M)$, we exploit the exact mapping
between the height field in the PNG model and the top path of $p$
non-intersecting 
random walkers, so called vicious walkers \cite{fisher}, in the limit
$p \to \infty$ 
\cite{praehofer_spohn, ferrari_beg}. Here we consider
``watermelons'' (Fig. \ref{fig_intro} c)) where $p$ non-colliding Brownian motions $x_1(\tau)<\cdots < x_p(\tau)$
 on the unit time interval are
constrained to start and end at $0$ ({\it i.e.} Brownian bridges). In
the large $p$ limit, one can show, using the connection between this
vicious walkers problem and random matrix theory, that $x_p(\tau)$  
also reaches a circular shape of amplitude $\sqrt{p}$, $x_p(\tau) \sim 2\sqrt{p}\sqrt{\tau(1-\tau)}$ while the fluctuations
are in that case of order $p^{-\frac{1}{6}}$ \cite{wt_prl}. Hence $x_p$ and $\tau$ map onto $h$
and $x$ in the growth model while $p$ plays the role of $t$. This mapping, for $p,t \gg 1$ reads~\cite{praehofer_spohn, ferrari_beg}
\begin{eqnarray}\label{def_airy}
 \frac{h(u t ^{\frac{2}{3}},t) - 2 t}{t^{\frac{1}{3}}} \equiv \frac{x_p(\frac{1}{2}+ \frac{u}{2}p^{-\frac{1}{3}}) - \sqrt{p}}{p^{-\frac{1}{6}}} \equiv {\cal A}_2(u) - u^2
\end{eqnarray}
where ${\cal A}_2(u)$ is the Airy$_2$ process \cite{praehofer_spohn} which is a stationary, and non-Markovian, process. In particular, 
${\rm Proba}[{\cal A}_2(0) \leq \xi] = {\cal F}_2(\xi)$. In this Letter, we compute exactly the joint distribution
$P_p(M,\tau_M)$ of the maximal height $M$ and its position $\tau_M$ for
the vicious walker problem (Fig. \ref{fig_intro} c)). While the maximal height $M$ has been recently
studied \cite{wt_prl, feierl2}, nothing is  
known about the distribution of $\tau_M$, which has by the way
recently attracted much interest in
various other one-dimensional stochastic processes 
\cite{satya_yor, dist_tmax, convex, hitting}. Our results are not only
relevant, for finite $p$, for the vicious walkers problem, but 
thanks to the above relation 
(\ref{def_airy}), become exact for $p \to \infty$, for curved growing
interface, as well as for the DPRM. We actually show that for moderate
values of $p \sim 10$, our analytical formula describes quite accurately our
numerical data for the PNG model.

\section{Method} The basic idea of the method to compute $P_p(M, \tau_M)$
is to divide the ''watermelons'' configuration in two time intervals, $\tau \in [0,\tau_M]$ and
$\tau \in [\tau_M, 1]$ and use the Markov property of the whole process
to treat these two intervals independently  (Fig. \ref{fig_intro} d)). In both  
intervals, the $p$ vicious walkers are constrained to stay below $M$, while we impose $x_p(\tau_M) = M$. 
To compute the propagator of these constrained vicious walkers in each sub-interval, we use a path-integral approach. {Let us denote 
$p_{<M}({\mathbf b},t_b | {\mathbf a},t_a)$ the  propagator of $p$ non-intersecting Brownian motions, 
starting in ${\mathbf a}\equiv (a_1, \cdots, a_p)$ at time $t_a$ and ending in ${\mathbf b} \equiv (b_1, \cdots, b_p)$ 
at time $t_b$  and constrained to stay below $M$ in the time interval $[t_a,t_b]$.  $p_{<M}({\mathbf b},t_b | {\mathbf a},t_a)$ is given by the sum
of the weights $\exp{\left[-\frac{1}{2}\sum_{i=1}^p \int_{t_a}^{t_b} \left(\frac{dx_i}{d\tau} \right)^2 \rmd \tau\right]}$ over all trajectories satisfying $x_1(\tau) < x_2(\tau) < \cdots < x_p(\tau) < M$, for $\tau \in [t_a,t_b]$. In the language of path integrals, this correponds to the propagator (in imaginary time) of 
$p$ quantum free fermions with an infinite wall in $x=M$, the associated Schr\"odinger Hamiltonian being $H_M = -\frac{1}{2} \sum_{i=1}^p \partial^2_{x_i} + V_M(x)$. The hard wall potential is given by $V_M(x) = 0$ if $x < M$ and $V_M(x) = + \infty$ if $x > M$ \cite{wt_prl}. The use of fermions incorporates naturally the non-colliding condition \cite{pgdg,wt_prl, celine_wishart}. This allows to write this propagator as
\begin{equation}\label{propag_1}
p_{<M}({\mathbf b},t_b |
{\mathbf a},t_a) = \langle {\mathbf b}|e^{-(t_b-t_a)H_M}| {\mathbf a}
\rangle \;,
\end{equation}
which we can compute using a spectral decomposition over the fermionic eigenfunctions of $H_M$.} Before doing this, we notice that the ''watermelons'' configurations that we study here are actually ill-defined for Brownian motions which are continuous both in space and time. {It is indeed well known that if two walkers
cross each other they will recross each inifinetly many times immediately after the first crossing}. This means in particular that it 
is impossible to impose $x_i(0) = x_{i+1}(0)$ and simultaneously $x_i(0^+) < x_{i+1}(0^+)$. Here, following Ref. \cite{satya_airy, wt_prl, satya_yor}, we adopt a regularization scheme where we impose
that the $p$ walkers start and end at $0 < \epsilon\, < \cdots < (p-1) \epsilon$ and 
take eventually the limit $\epsilon \to 0$. We use an additional cut-off procedure by
imposing that $x_p(\tau_M) = M - \eta$ and then take the limit $\eta \to 0$. 

\section{Results for $p$ vicious walkers} 

The calculation of $P_p(M,\tau_M)$ requires the computation of $p_{<M}({\mathbf b},t_b | {\mathbf a},t_a)$. {Expanding Eq. (\ref{propag_1}) over the fermionic eigenvectors of $H_M$ yields}
\begin{multline}
\label{propag}
p_{<M}({\mathbf b},t_b |
{\mathbf a},t_a) = \langle {\mathbf b}|e^{-(t_b-t_a)H_M}| {\mathbf a}
\rangle \\
 = \int_{0}^\infty \rmd {\mathbf k}\, e^{-\frac{{\mathbf k}^2}{2}(t_b-t_a)}
\det_{1 \leq i,j \leq p} \phi_{k_i}(b_j) \det_{1 \leq i,j \leq p}
\phi^*_{k_i}(a_j)\;, 
\end{multline}  
where $\phi_k(x) = \sqrt{\frac{2}{\pi}}
\sin{\left[ k(M-x) \right]}$ naturally appear as the eigenvectors of
$H_M$ and
where we use the notations $\int_{0}^\infty \rmd {\mathbf k} \equiv
\int_{0}^\infty dk_1 \cdots \int_{0}^\infty dk_p$ and
${\mathbf k}^2 = k_1^2 + \cdots + k_p^2$. In Eq. (\ref{propag}), the
determinants appear as Slater determinants in the 
associated fermions problem. From this propagator (\ref{propag}) we compute 
$P_p(M,\tau_M)$ by dividing the configuration in two time independent intervals, $\tau \in [0,\tau_M]$ and
$\tau \in [\tau_M, 1]$ as explained above (see also Fig. \ref{fig_intro} d))  to obtain:
\begin{multline}
\label{start}
P_p(M, \tau_M) = \lim_{{\epsilon}, \eta \to 0} \frac{1}{Z_p}
\int\limits^{M-\eta}_{-\infty} \!\! \rmd {\mathbf y} \,  p_{<M}(\gras{\epsilon},1 |
{\mathbf y},\tau_M) \\
\times p_{<M}({\mathbf y},\tau_M | \gras{\epsilon},0) \delta(y_p - (M-\eta)) \;,
\end{multline}
where the delta function enforces $x_p(\tau_M) = M-\eta$ and where the amplitude
$Z_p$, which depends explicitly on 
$\epsilon$ and $\eta$ is determined by the normalization condition
$\int_{0}^{+\infty} \rmd M \int_0^1 \rmd \tau_M P_p(M, \tau_M) =
1$. Using the
above formula for the constrained propagator (\ref{propag}) in Eq. (\ref{start}), and taking the limits $\epsilon, \eta \to 0$ one
obtains
\begin{multline}\label{expr_inter}
P_{p}(M, \tau_M) = {z^{-1}_p} M^{-(p^2+3)}
\int_{0}^\infty \rmd {\mathbf q} \int_{0}^\infty \rmd q'_p
e^{- \frac{\sum_{i=1}^{p-1} q_i^2}{2M^2}}  \\
\times\, e^{ - \frac{\tau q_p^2 +(1-\tau){q'_p}^2}{2M^2}  }  q_p q'_p
\Theta_p(q_1, \cdots, q_p)\Theta_p(q_1, \cdots, q'_p) ,
\end{multline} 
where $\Theta_p({\mathbf q})$ is the following determinant
\begin{eqnarray}
\Theta_p({\mathbf q}) = \det_{1\leq i,j \leq p} q_i^{j-1} \cos{\left(q_i - j {\pi}/{2}\right)} \;.
\end{eqnarray}
To compute $z_p$, we use that $\int_0^1 P_p(M,\tau_M) \rmd
\tau_M$ must yield back the expression for the distribution of the
maximum as computed in Ref. \cite{wt_prl}. This allows to obtain $z_p =  \pi^{1+\frac{p}{2}} 2^{-\frac{3p}{2}} \prod_{j=0}^{p-1} j!$.
In Eq. (\ref{expr_inter}), one can expand the determinant by
minors and then 
perform the integrals over $q_i$ using the Cauchy-Binet
identity which reads
\begin{multline}
\int \rmd \gras{x} \det_{1\leq i,j \leq p} f_i(x_j) \det_{1\leq i,j \leq p} g_i(x_j) \\= p! \det_{1\leq i,j\leq p} \int \rmd x f_i(x) g_j(x),
\end{multline}
for any suitable functions $f_i(x)$ and $g_i(x)$.
These integrals can be expressed in terms of Hermite
polynomials $H_n(x)$ and we obtain finally
\begin{equation}\label{determinantal_expr}
P_p(M,\tau_M) = B_p[\det \mathrm{D}]~^t\mathrm{U}(\tau_M)  \mathrm{D}^{-1} \mathrm{U}(1-\tau_M) \;,
\end{equation}
with $B_p^{-1} = \left( 2 \pi \right)^{\frac{1}{2}}\prod_{j=0}^{p-1} \left( j! 2^j \right)$ and 
where $\mathrm{D}\equiv \mathrm{D}(M)$ is a $p \times p$ matrix 
\begin{equation}
\label{matrix_D}
\mathrm{D}_{i,j}= (-1)^{i-1} H_{i+j-2}(0) - e^{-2M^2} H_{i+j-2}(\sqrt{2}M) \;,
\end{equation}
while $\mathrm{U(\tau)} \equiv \mathrm{U}(M,\tau)$ is a column vector given by 
\begin{equation}
\mathrm{U}_i(\tau) =\tau^{-\frac{i+1}{2}} \ H_i\left({M}/{\sqrt{2 \tau}}\right)\  e^{-\frac{M^2}{2 \tau}} .
\end{equation}

\begin{figure}[h]
\onefigure[width=  .5\linewidth]{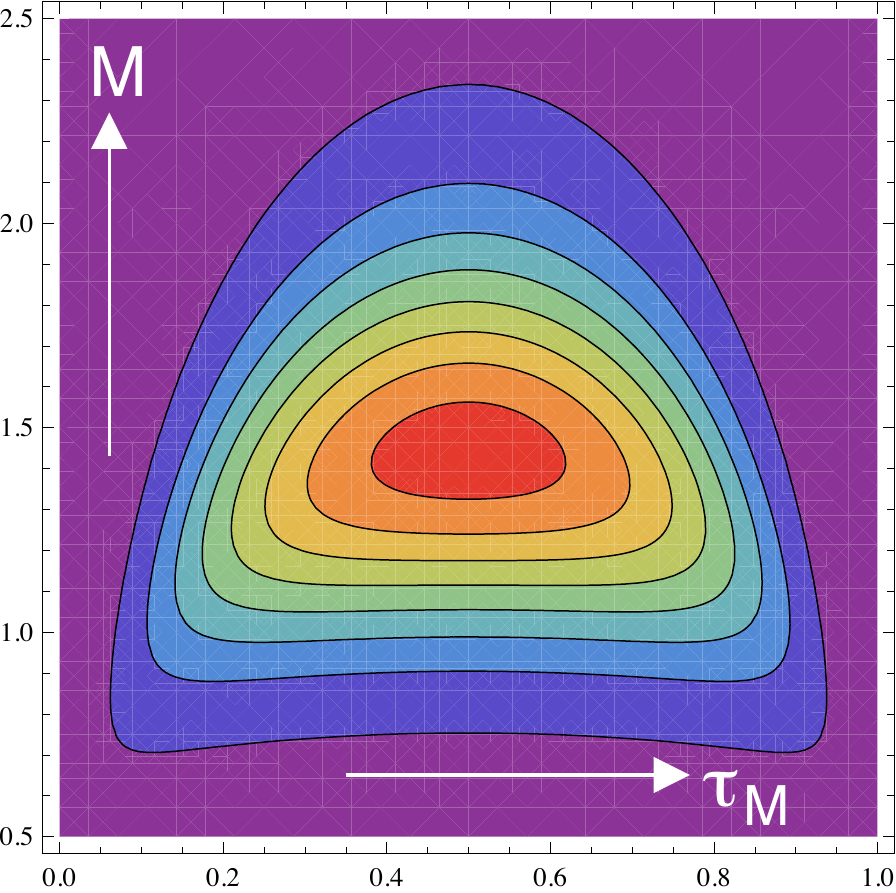}
\caption{Contour plot of $P_p(M,\tau_M)$ for $p=3$. The contour
lines correspond to $0.1, 0.4, \cdots, 2.2$.}
\label{contourplot}
\end{figure}

In Fig. (\ref{contourplot}), we show a contour plot  of
$P_p(M,\tau_M)$ for $p=3$. For fixed $\tau_M \in [0,1]$,
$P_p(M,\tau_M)$, as a function of $M$ 
has a simple bell shape. Its behavior for fixed value of $M$ as a
function of $\tau_M$ 
is more interesting. For sufficiently large $M > M^*_p$, it has a bell shape, with a maximum in $\tau_M=\frac{1}{2}$, while for $M < M^*_p$, it
has a ''M-shape'' with two distinct maxima, $\tau_M=\frac{1}{2}$ being a local minimum. One observes that $M^*_p$ is a
slowly increasing function of $p$.

By integrating our expression in Eq. (\ref{determinantal_expr}) over $\tau_M$, one checks  
that we recover the formula for the pdf of $M$, as obtained in Ref. \cite{feierl2}. Indeed, one finds ${\rm Proba}[\max_{\tau} x_p(\tau) \leq M] = 
\det {\mathrm{D}}/\prod_{j=0}^{p-1} \left( j! 2^j \right)$, where the matrix $\mathrm{D}$ is defined in Eq.~(\ref{matrix_D}). 
On the other hand, by integrating Eq. (\ref{determinantal_expr}) over $M$, 
one obtains an expression for $P_p(\tau_M)$. While for $p=1$, $P_1(\tau_M)=1$, one obtains, for instance, for $p=2$
\begin{eqnarray}
P_2(\tau_M) = 4 \left(1 - \frac{1+10 \tau_M(1-\tau_M)}{(1+4\tau_M(1-\tau_M))^{5/2}} \right) \;. 
\end{eqnarray}
For generic $p$, $P_p(\tau_M)$ is a function of
$\tau_M(1-\tau_M)$ with the asymptotic behavior, for small $\tau_M$
\begin{equation}
P_p(\tau_M) \sim \tau_M^{\nu(p)},\quad \text{with $\nu(p) = (p^2+p-2)/{2}$.}
\end{equation} 
Note that for $p$
independent Brownian 
motions (without the non-crossing condition) one has $P_{p,{\mathrm{free}}}(\tau_M) \sim \tau_M^{p-1}$ \cite{convex} so that the exponent $\nu(p)$ bears the signature of the non-colliding condition. 

The above approach can be extended to study the extreme statistics of
$p$ non-intersecting excursions, {\it i.e.} vicious walkers starting and
terminating at the origin but with the additional constraint that they all
stay positive 
inbetween \cite{katori_p2, wt_prl, feierl1}. The slight modification in our computation is to replace the Hamiltonian $H_M$
(with a wall in $x=M$) with the box Hamiltonian $H_{\text{Box}}$ (with two walls: in $x=0$ and in $x=M$).
Hence the energy levels are discrete, and consequently one has to treat discrete sums instead of integrals as before.
Using the appropriate propagator in Eq.~(\ref{start}), one finds, taking the limits $\epsilon, \eta \to 0$, 
the joint pdf for the non-intersecting excursions
\begin{multline}
\label{excursion_intermediaire}
P_{p,E}(M,\tau_M)=z_{p,E}^{-1}\ M^{-(2p^2+p+3)}\\
\sum_{n_1,\dots,n_p,n'_p >0} (-1)^{n_p+n'_p}\
e^{-\frac{\pi^2}{2M^2}\left[ \sum_{i=1}^{p-1} n_i^2 + \tau_M n_p^2 +(1-\tau_M){n'_p}^2\right] }
\\
\times \prod_{i=1}^{p-1} n_i^2 \ n_p^2 \ {n'_p}^2\ \Delta_p(n_1^2,\dots,n_p^2) \Delta_p(n_1^2,\dots,{n'_p}^2) ,
\end{multline}
where we use the notation $\Delta_p(\lambda_1,\dots,\lambda_p)=\prod_{1\leq i <j \leq p} (\lambda_j-\lambda_i)$
for the Vandermonde determinant ,
and with the normalization $z_{p,E}=2^{\frac{p^2}{2}} \Gamma(p) \prod_{j=0}^{p-1} (2^j j! \Gamma(\frac{3}{2}+j)) \pi^{-(2p^2+p+2)}$.
Performing similar manipulations as before, 
one finds the joint pdf of the couple $(M,\tau_M)$ for the excursions configuration
in a determinantal form, reminiscent of the watermelons case~(\ref{determinantal_expr}):
\begin{equation}\label{determinantal_expr_excursions}
P_{p,E}(M,\tau_M) = C_p[\det \mathrm{D}_E]~^t\mathrm{U}_E(\tau_M)  \mathrm{D}_E^{-1} \mathrm{U}_E(1-\tau_M) \;,
\end{equation}
with $C_p^{-1}=(-1)^{p+1}2^{2p^2-\frac{1}{2}} \prod_{j=1}^{p-1} (j! \Gamma(\frac{3}{2}+j)) \pi^{-1} $.
The $p\times p$ matrix $\mathrm{D}_E \equiv \mathrm{D}_E(M)$ is (for $1\leq i,j \leq p$)
\begin{equation}
{\mathrm{D}_E}_{i,j}=\sum_{n=-\infty}^{+\infty} H_{2(i+j-1)}(\sqrt{2}M n) e^{-2 M^2 n^2},
\end{equation}
and appears in the expression of the cumulative distribution of the maximum~\cite{katori_p2}, 
$\mathrm{Proba}[\max_{\tau} x_p(\tau) \leq M] = (-1)^p\det \mathrm{D}_E/(2^{p^2} \prod_{j=1}^{p}(2j-1)!)$.
The vector elements are (for $i=1,\dots,p$)
\begin{equation}
{\mathrm{U}_E}_i(\tau)=\frac{1}{M} \left(-\frac{2 \pi^2}{M^2} \right)^i \sum_{n=1}^{\infty} (-1)^n n^{2i} e^{-\frac{2 \pi^2}{M^2} n^2 \tau} .
\end{equation}
The integration of formula~(\ref{excursion_intermediaire}) with respect to $M$ gives the pdf of the time to reach the maximum
in this excursion process.
For $p=1$, the formula in Eq.~(\ref{excursion_intermediaire}) yields back the result obtained in Ref.~\cite{satya_yor}.
For $p=2$, we give the expression of $P_{2,E}(\tau_M)$
\begin{equation}
P_{2,E}(\tau_M) = a \sum_{n_i>0}  \frac{(-1)^{n_2+n_3}\, n_1^2 n_2^2 n_3^2(n_1^2-n_2^2)(n_1^2-n_3^2)}{(n_1^2 + \tau_M n_2^2 + (1-\tau_M)n_3^2)^6} ,
\end{equation}
with $a = 1280/\pi$.

\section{Application to stochastic growth processes} We now come back to the joint pdf $P_t(M,X_M)$ for 
curved growing interfaces (Fig.~\ref{fig_intro} a)). From Eq.~(\ref{def_airy}) one obtains
\begin{eqnarray}\label{recall_mapping}
 P_t(M,X_M) \sim t^{-1} {\cal P}_{{\mathrm{Airy}}} ((M-2t)t^{-\frac{1}{3}},X_Mt^{-\frac{2}{3}})
\end{eqnarray}
where ${\cal P}_{{\mathrm{Airy}}} (y,x)$ is the joint distribution of
the maximum $y$ and its position $x$ for the process ${\cal
  A}_2(u)-u^2$, $u \in \mathbb{R}$. From Ref. \cite{krug_dprm, baik_rains, johansson_transverse} one 
obtains that the (marginal) pdf of $y$, ${\cal P}_{{\mathrm{Airy}},M}(y)$, is given by TW, ${\cal P}_{{\mathrm{Airy}},M} (y) = {\cal F}'_1(y)$, and
hence $P_t(M) \sim t^{-\frac{1}{3}} {\cal F}'_1((M-2t)/t^{\frac{1}{3}})$. On the other hand, from Eq. (\ref{def_airy}), ${\cal P}_{{\mathrm{Airy}}} (y,x)$ can be obtained from our results for $P_p(M, \tau_M)$
in Eq. (\ref{determinantal_expr}): in the scaling limit $y = (M-\sqrt{p})p^{\frac{1}{6}}$ as well as $x = 2(\tau_M-\frac{1}{2})p^{\frac{1}{3}}$ fixed one has indeed 
$ P_p(M,\tau_M) \sim 2p^{\frac{1}{2}} {\cal P}_{{\mathrm{Airy}}}
(y,x)$. While our expression
(\ref{determinantal_expr}) should be amenable to an asymptotic analysis
for large $p$, yielding an explicit exact expression for ${\cal
  P}_{{\mathrm{Airy}}} (y,x)$, such an analysis deserves further investigations. Here we show instead that, for finite values of $p$, this expression (\ref{determinantal_expr}) describes quite accurately our numerical
data for the extreme statistics of the PNG model in the 
droplet geometry. To illustrate this, we have computed numerically the
(marginal) distribution of the position of the maximum $X_M$. We
recall that, for the DPRM as in Fig. (\ref{fig_intro} b)), $X_M$ is the position
of the free end of the optimal polymer. 

\begin{figure}[h]
\onefigure[width=\linewidth]{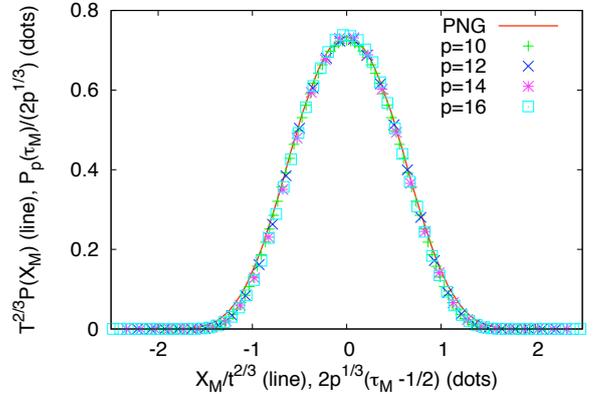} 
\caption{On linear scales, the solid line is a plot of $t^{\frac{2}{3}}P_t(X_M)$ as a function of $X_M/t^{\frac{2}{3}}$, for $t=768$,
computed numerically for the PNG, while the dots correspond to our analytical results $P_p(\tau_M)(2p^{\frac{1}{3}})$ 
(\ref{determinantal_expr}) as a function of $2p^{\frac{1}{3}}(\tau_M-\frac{1}{2})$ for $p=10, 12, 14$ and $p=16$.}
\label{scaling_png}
\end{figure}

In Fig. \ref{scaling_png},
we show a plot of the rescaled distribution $t^{\frac{2}{3}}P_t(X_M/t^{\frac{2}{3}})$
as a function of the rescaled variable $X_M/t^{\frac{2}{3}}$ for $t = 768$ in
solid line. We also plot our exact analytical results for watermelons,
{\it i.e.} $P_p(\tau_M)/(2 p^{\frac{1}{3}})$ as a function of
$(2p^{\frac{1}{3}}(\tau_M-\frac{1}{2}))$ for $p=10, 12, 14$ and $16$ which is computed from
Eq. (\ref{determinantal_expr}) as $P_p(\tau_M) = \int_0^\infty dM
P_p(M,\tau_M)$. We emphasize that the good collapse of the different
curves is obtained without any fitting parameter.

\begin{figure}[h]
\onefigure[width=\linewidth]{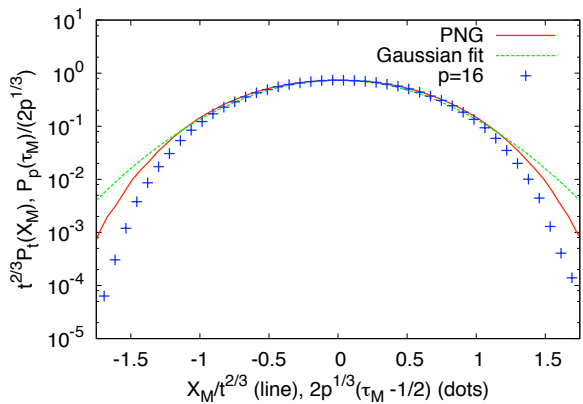}
\caption{Log-linear plot of: 
the rescaled distribution of the position of the maximum for PNG (solid line),
the rescaled distribution of the time to reach the maximum for the $p=16$ watermelons configuration (dots),
and the closest Gaussian curve (dashed line), that do not match the PNG curve.
The scaled quantities appears in the $x,y$ labels 
(and are detailed in the caption of Fig~\ref{scaling_png}).}
\label{gaussian_png}
\end{figure}

In Fig.~\ref{gaussian_png}, we show a plot of the same quantities as in
Fig.~\ref{scaling_png} in log-linear plot. As expected, one observes
some deviations between the result for the PNG and our computation for
finite $p=16$ in the tail of the distribution. Besides, our numerical
data suggest that the marginal distribution of $X_M$ is {\it
  non-Gaussian} (the best Gaussian approximation being shown as a
dotted line). Instead, for $X_M \gg t^{2/3}$, our data are compatible
with a stretched 
exponential behavior $P_t(X_M) \sim t^{-\frac{2}{3}} e^{-\gamma
  \left(\frac{X_M}{t^{2/3}}\right)^\delta}$ with $\delta \simeq 2.5$
while its precise determination requires more numerical efforts. Interestingly, for $p$ free Brownian bridges, one
can show \cite{convex} that the distribution of $\tau_M$ converges to
a Gaussian 
distribution centered in $\frac{1}{2}$ of 
width $(8 \log {n})^{-\frac{1}{2}}$. Therefore, the non-Gaussianity of $P_t(X_M)$ is a clear
signature of the correlations in the associated vicious walkers problem. 

{The results obtained in the present Letter extend far beyond the PNG model. Indeed our results hold more generally for stochastic
growth models, and physical situations, where the interface $h(x,t)$ at point $x$ and time $t$ evolves
according to the one-dimensional KPZ equation \cite{kpz}
\begin{eqnarray}\label{KPZ}
\frac{\partial}{\partial t} h(x,t) = \nu \nabla^2 h(x,t) + \frac{\lambda}{2}(\nabla h(x,t))^2 + \zeta(x,t) \;,
\end{eqnarray}
in a curved geometry. In Eq. (\ref{KPZ}), $\zeta(x,t)$ is a Gaussian white noise of zero mean and correlations $\langle \zeta(x,t) \zeta(x',t') \rangle = D \delta(x-x') \delta(t-t')$. In Ref. \cite{praehofer_spohn}, it was shown that the fluctuations of physical obervables of the height field, evolving according to Eq. (\ref{KPZ}), are universal, up to two non-universal parameters $\lambda$ and $A = D/2\nu$. Once $\lambda$ and $A$ are fixed, these fluctuations are characterized by universal  distribution functions. In the curved geometry, $\lambda$ can be simply measured as the radial growth rate, $\langle h(x=0,t) \rangle \sim \lambda t$ while $A$ can be extracted from the width of the interface $\langle (h(0,t) - \langle h(0,t)\rangle)^2 \rangle^{1/2} \sim A L/6$ for a system of finite size $L$ \cite{krug_dprm}. For instance, the fluctuations of the height field are given by $h(0,t) \simeq \lambda t + (A^2 \lambda t/2)^{1/3} \chi_{\rm GUE}$ where $\chi_{\rm GUE}$ is a random variable distributed according to the TW distribution ${\cal F}_2$ \cite{praehofer_spohn}. This universality was convincingly demonstrated in recent experiments \cite{exp_curved}. Therefore, for an interface described by the KPZ equation (\ref{KPZ}) in a droplet geometry, one expects that the distribution of the position of the maximum $X_M$ will have the scaling form $P_t(X_M) \sim \xi(t)^{-1} {\cal P}_{{\rm Airy},X_M}(X_M/\xi(t))$ with $\xi(t) = (2/A) ({A^2 \lambda t}/{2})^{2/3}$ is the correlation length. The scaling function ${\cal P}_{{\rm Airy},X_M}(x)$ is universal and can be computed from our formulas in Eq. (\ref{determinantal_expr}, \ref{recall_mapping}), see also Fig. \ref{scaling_png}.  

}

\section{Conclusion} To conclude, we have obtained an exact expression
for the joint distribution of the maximum $M$ and the time $\tau_M$ at
which this maximum is reached for $p$ non-intersecting Brownian
bridges (\ref{determinantal_expr}) and excursions~(\ref{determinantal_expr_excursions}). 
We have shown that our analytic
expression for moderate values 
of $p$ Brownian bridges describe very accurately the extreme statistics of curved
growing interfaces, becoming eventually exact in the limit $p \to
\infty$. In addition to their relevance to the DPRM, our results may
have applications to 
various other situations like step fluctuations in faceted crystals \cite{ferrari} or
dimer-covering problems \cite{johansson_arctic}, where it was shown that the fluctuations 
are governed by the very same process ${\cal A}_2(u) - u^2$ that we
have studied here. Finally, in view of recent progresses
\cite{exp_planar, exp_curved}, it seems possible to observe 
these extremal statistics for curved growing interfaces in 
experimental situations like nematic liquid crystals \cite{exp_curved}.    
 
\acknowledgments
We thank T. Sasamoto for useful correspondence.

\end{document}